\documentclass[5p,sort&compress]{elsarticle}
\usepackage{epstopdf}
\usepackage{cuted}
\usepackage{subfigure,graphicx}
\usepackage{float}
\usepackage{amsmath, amsfonts, amssymb}
\usepackage{amsthm}
\usepackage{epsf}
\usepackage{amsfonts}
\usepackage{stmaryrd}
\usepackage{amssymb}
\usepackage{leftidx}
\usepackage{color}
\usepackage{mathtools}
\usepackage{placeins}
\usepackage{booktabs}
\usepackage{enumitem}
\usepackage{caption}
\usepackage{tabu}
\usepackage{multirow}
\usepackage{stackengine}
\usepackage[utf8]{inputenc}
\usepackage[english]{babel}
\usepackage{color}

\theoremstyle{plain}
\newtheorem{theorem}{Theorem}

\newtheorem{remark}[theorem]{Remark}

\def\bfx{{\bf x}}
\def\bfy{{\bf y}}

\def\bfC{{\bf C}}

\def\bfI{{\bf I}}

\def\bfN{{\bf N}}

\def\bfS{{\bf S}}

\def\bfX{{\bf X}}
\def\bfF{{\bf F}}

\def\bfe{{\bf e}}

\def\e0{\varepsilon_0}

\def\s0{\sigma_0}

\long\def\symbolfootnote[#1]#2{\begingroup%
\def\thefootnote{\fnsymbol{footnote}}\footnote[#1]{#2}\endgroup}

\begin{document}
\begin{frontmatter}

\title{The trousers fracture test for viscoelastic elastomers}

\author{Bhavesh Shrimali}
\ead{bshrima2@illinois.edu}

\author{Oscar Lopez-Pamies}
\ead{pamies@illinois.edu}

\address{Department of Civil and Environmental Engineering, University of Illinois, Urbana--Champaign, IL 61801, USA  \vspace{0.2cm}}

\vspace{0.9cm}

\begin{abstract}

\vspace{0.25cm}

Shrimali and Lopez-Pamies (2023) have recently shown that the Griffith criticality condition that governs crack growth in viscoelastic elastomers can be reduced to a fundamental form that involves exclusively the intrinsic fracture energy $G_c$ of the elastomer and, in so doing, they have brought resolution to the complete description of the historically elusive notion of critical tearing energy $T_c$. The purpose of this paper --- which can be viewed as the third installment of the series started by Shrimali and Lopez-Pamies (2023) --- is to make use of this fundamental form to explain one of the most popular fracture tests for probing the growth of cracks in viscoelastic elastomers, the trousers test.

\vspace{0.3cm}

\endkeyword

\end{abstract}

\end{frontmatter}

\section{Introduction}\label{Sec:IntroMain}

In a recent contribution, Shrimali and Lopez-Pamies (2023a) have shown that the original form (Rivlin and Thomas, 1953; Greensmith and Thomas, 1955)
\begin{equation}
-\dfrac{\partial \mathcal{W}}{\partial \mathrm{\Gamma}_0}=T_{c}\label{Tc-0}
\end{equation}
of the Griffith criticality condition that describes the growth of cracks in elastomers subjected to quasi-static mechanical loads can be reduced to the fundamental form
\begin{equation}
-\dfrac{\partial \mathcal{W}^{{\rm Eq}}}{\partial \mathrm{\Gamma}_0}=G_{c}.\label{Gc-0}
\end{equation}
In expression (\ref{Tc-0}), the left-hand side $-\partial \mathcal{W}/\partial \mathrm{\Gamma}_0$ denotes the change in total deformation (stored and dissipated) energy $\mathcal{W}$ in the bulk with respect to an added surface area to the pre-existing crack $\Gamma_0$, while the right-hand side $T_c$ stands for the critical tearing energy, \emph{a characteristic property of the elastomer that depends on the loading history}. In expression (\ref{Gc-0}), on the other hand, $-\partial \mathcal{W}^{{\rm Eq}}/\partial \mathrm{\Gamma}_0$ denotes the change in equilibrium elastic energy $\mathcal{W}^{{\rm Eq}}$ stored in the bulk with respect to an added surface area to the pre-existing crack $\Gamma_0$, while $G_c$ stands for the intrinsic fracture energy, \emph{a material constant of the elastomer}. Experiments have shown that its value is typically in the same relatively narrow range
\begin{equation}
G_{c}\in[10,100]\, {\rm N}/{\rm m}\label{Gc-100}
\end{equation}
for many common elastomers (Ahagon and Gent, 1975; Gent and Tobias, 1982; Bhowmick et al., 1983).

Historically, the shortcoming of the criticality condition (\ref{Tc-0}) has been \emph{not} knowing how the critical tearing energy $T_c$ depends on the applied loading history for a given elastomer and given geometry of the body of interest. The work of Shrimali and Lopez-Pamies (2023a) has brought resolution to this decades-old problem by first recognizing that the total deformation energy $\mathcal{W}$ admits the partition
\begin{equation}
\mathcal{W}=\underbrace{\mathcal{W}^{{\rm Eq}}+\mathcal{W}^{{\rm NEq}}}_\text{stored}+\underbrace{\mathcal{W}^{v}}_\text{dissipated} \label{WWW}
\end{equation}
for any viscoelastic elastomer\footnote{The interested reader is referred to (Kumar and Lopez-Pamies, 2016) for a detailed account on the viscoelasticy of elastomers. Here, we merely recall that rheological representations of elastomers provide a helpful visualization of the energy partition (\ref{WWW}). For instance, in the Zener-type rheological representation depicted in Fig. \ref{Fig2}, $\mathcal{W}^{{\rm Eq}}$ and $\mathcal{W}^{{\rm NEq}}$ correspond to the elastic energy stored in the equilibrium and non-equilibrium springs, whereas $\mathcal{W}^{v}$ corresponds to the viscous energy dissipated by the dashpot.} and then establishing from experiments that
\begin{equation}
T_c=G_c-\dfrac{\partial \mathcal{W}^{{\rm NEq}}}{\partial \mathrm{\Gamma}_0}-\dfrac{\partial \mathcal{W}^{v}}{\partial \mathrm{\Gamma}_0}\label{Tc-for}
\end{equation}
at fracture. In these expressions, wherein dissipation mechanisms other than viscous deformation (e.g., strain-induced crystallization) are assumed absent, $\mathcal{W}^{v}$ represents the part of the total energy that is dissipated by the elastomer via viscous deformation, while the combination $\mathcal{W}^{{\rm Eq}}+\mathcal{W}^{{\rm NEq}}$ represents the part of the total energy that is stored by the elastomer via elastic deformation. Precisely, $\mathcal{W}^{{\rm NEq}}$ stands for the part of the stored elastic energy that will be dissipated eventually via viscous dissipation as the elastomer reaches a state of thermodynamic equilibrium. On the contrary, $\mathcal{W}^{{\rm Eq}}$ denotes the part of the stored elastic energy that the elastomer will retain at thermodynamic equilibrium.

From a fundamental point of view, the criticality condition (\ref{Gc-0}) is a strikingly simple and intuitive condition as it states that whether an elastomer simply deforms or, on the other hand, creates new surface from a pre-existing crack is dictated by a competition between its stored equilibrium elastic energy and its intrinsic fracture energy, irrespective of its viscosity.

From a practical point of view, the criticality condition (\ref{Gc-0}) is also conveniently simple. This is because it is based on two properties of the elastomer that can be measured experimentally once and for all by means of conventional tests: (\emph{i}) its viscoelastic behavior, from which the storage of equilibrium elastic energy can be identified, and (\emph{ii}) its intrinsic fracture energy.

As a first effort to gain precise and quantitative insight into the fracture behavior of viscoelastic elastomers, Shrimali and Lopez-Pamies (2023a,b) have made use of the newly-minted fundamental form (\ref{Gc-0}) of the Griffith criticality condition to explain two types of popular fracture tests for viscoelastic elastomers, the so-called ``pure-shear'' fracture test and the delayed fracture test. The object of this paper --- which can be viewed as the third instalment of the series --- is to deploy the criticality condition (\ref{Gc-0}) to explain yet another popular fracture test: the trousers test.

The paper is organized as follows. In the same spirit of the pioneering global (elastic) analysis provided by Rivlin and Thomas (1953), we begin in the first part of Section \ref{Sec:Global} by presenting a global analysis of fracture nucleation in the trousers test for viscoelastic elastomers. In the same spirit of the global analysis provided by Greensmith and Thomas (1955), we also present in the latter part of Section \ref{Sec:Global} a global analysis of fracture propagation. In Section \ref{Sec:Full-field}, in preparation to carry out corresponding full-field analyses, we formulate the initial-boundary-value problem of the trousers test \emph{per se}. In Section \ref{Results Gaussian}, we present and discuss full-field solutions for the canonical case of a viscoelastic elastomer with Gaussian elasticity and constant viscosity. These results serve to lay bare the key features of the trousers test in the simplest of settings. We conclude in Section \ref{Sec: Final Comments} by summarizing the main findings of this work and by recording a number of closing remarks.

\section{Global analysis of the trousers test}\label{Sec:Global}

\subsection{Fracture nucleation}

\begin{figure}[t!]\centering
 \includegraphics[width=3.4in]{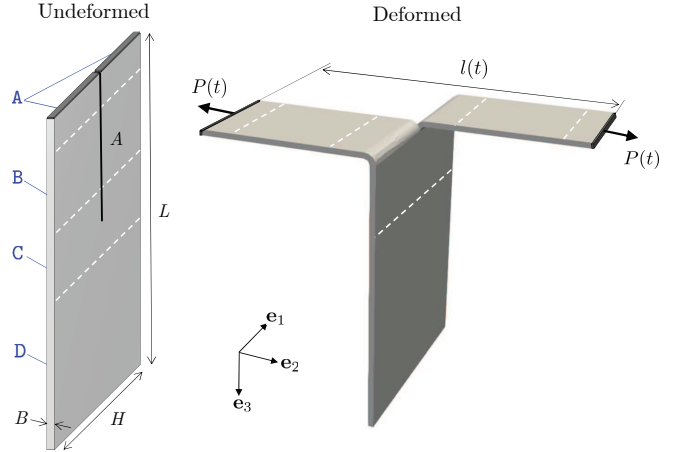}
\caption{\small Schematic of the trousers test for a viscoelastic elastomer. The dimensions in the undeformed configuration are such that $B\ll H< A < L$. The two bottom ends of the ``trousers'' are held firmly by stiff grips and then pulled apart either by an applied force $P(t)$ or by an applied deformation $l(t)$. Provided that the deformation $l(t)$ is not too large, the region $\texttt{B}$ in the specimen is essentially in a state of spatially uniform uniaxial tension. For this reason, this test is also sometimes referred to as a ``simple-extension'' fracture test.}
   \label{Fig1}
\end{figure}

Consider the trousers test schematically depicted in Fig. \ref{Fig1}. The specimen, which is taken to be made of an isotropic incompressible viscoelastic elastomer, is such that  its thickness is much smaller than its height ($B\ll H$), its height is smaller than its length ($H<L$), and the initial length of the pre-existing crack is larger than the height of the specimen but, obviously, smaller than its length ($H< A< L$). In experiments, one typically encounters the ranges of ratios $L/H\in[2,4]$,  $A/H \in[1,2]$, and $H/B\in[20,40]$.

The two legs of the trousers are bent opposite to one another and brought to lie in the same plane, their bottom ends are held firmly by stiff grips, and these are then pulled apart. In experiments, the pulling is done either by applying a force $P(t)$ or by applying a deformation $l(t)$ over a time interval $t\in[0,T]$; see, e.g., Fig. 2 in (Greensmith and Thomas, 1955). For the former case, a force is typically ramped up over an initial time interval $[0,t_0]$ and then held constant. When a deformation is applied, on the other hand, the grips are typically separated at a constant rate $\dot{l}_0$ so that the current distance between the grips is given by the relation $l(t)=l_0+\dot{l}_0 t$, where $l_0\approx 2A$ since $B\ll A$.

Because of the special geometry of the specimen, at any given time $t\in(0,T]$, there are four different regions of deformation; see Fig. \ref{Fig1}. In keeping with the same type of region labeling used by Rivlin and Thomas (1953) --- see Fig. 7 in their work --- the region $\texttt{D}$ is substantially undeformed, the crack-front region $\texttt{C}$ and the grip region $\texttt{A}$ are in a complex state of deformation (highly non-uniform in space), while region $\texttt{B}$ is substantially in a state of spatially uniform uniaxial tension, this provided that the deformation $l(t)$ between the grips is not exceedingly large. Large deformations $l(t)$ lead to significant twisting of the specimen, which in turn results in region $\texttt{B}$ also exhibiting a state of deformation that is \emph{not} uniform. In the remaining of this section, we tacitly assume that $l(t)$ is small enough and hence that the state of deformation in region $\texttt{B}$ is one of uniform uniaxial tension.

Now, for the case when the test is carried out by applying a force $P(t)$, consider an increase in the crack surface of amount ${\rm d}\mathrm{\Gamma}_0=B {\rm d}A$ at time $t$. This increase in crack surface does \emph{not} alter the complex states of deformation in $\texttt{A}$ and $\texttt{C}$. Instead, it simply shifts the region $\texttt{C}$ in the direction of the added crack, resulting in the growth of region $\texttt{B}$ at the expense of region $\texttt{D}$. In other words, an added crack ${\rm d}\Gamma_0$ at constant applied force $P(t)$ results in the transferring of a volume $H{\rm d}\Gamma_0$ of the specimen from an undeformed state to a state of uniaxial tension under the same force $P(t)$. Making use of this observation, we can immediately deduce that
\begin{equation}
\left.\dfrac{\partial \mathcal{W}^{{\rm Eq}}}{\partial \mathrm{\Gamma}_0}\right|_{P}=H \psi_{ut}^{{\rm Eq}}(\lambda(t)),\label{ERR-P}
\end{equation}
where the suffix $P$ denotes differentiation at fixed force $P(t)$, $\lambda(t)$ is the stretch that results by subjecting the elastomer to uniform uniaxial tension with a force $P(t)$, and $\psi_{ut}^{{\rm Eq}}$ stands for the equilibrium elastic energy density stored at that state of deformation.

By definition, the derivative in the Griffith criticality condition (\ref{Gc-0}) is to be taken at fixed deformation $l(t)$, and \emph{not} at fixed force $P(t)$. It follows from a result of Shrimali and Lopez-Pamies (2023b) that these two derivatives are related to one another according to the equality
\begin{equation}
\left.-\dfrac{\partial \mathcal{W}^{{\rm Eq}}}{\partial \mathrm{\Gamma}_0}\right|_{l}=P^{\rm{Eq}}\left.\dfrac{\partial l}{\partial \mathrm{\Gamma}_0}\right|_{P}-\left.\dfrac{\partial \mathcal{W}^{{\rm Eq}}}{\partial \mathrm{\Gamma}_0}\right|_{P}\label{ERR-l0}
\end{equation}
with
\begin{equation}
P^{\rm{Eq}}:=\left.\dfrac{\partial \mathcal{W}^{{\rm Eq}}}{\partial l}\right|_{\mathrm{\Gamma}_0}.\label{PEq}
\end{equation}
Upon recognizing from the geometry of the test that
\begin{equation*}
\left.\dfrac{\partial l}{\partial \mathrm{\Gamma}_0}\right|_{P}=\dfrac{2}{B}\lambda(t)
\end{equation*}
and making direct use of the result (\ref{ERR-P}), it follows from (\ref{ERR-l0}) that
\begin{equation}
-\dfrac{\partial \mathcal{W}^{{\rm Eq}}}{\partial \mathrm{\Gamma}_0}=\dfrac{2}{B} P^{\rm{Eq}}\lambda(t)-H \psi_{ut}^{{\rm Eq}}(\lambda(t)),\label{ERR-l}
\end{equation}
where we have reverted back to omitting the suffix $l$ in the derivative $-\partial \mathcal{W}^{{\rm Eq}}/$ $\partial \mathrm{\Gamma}_0$, since there is no longer risk of confusion.

This last relation is the result that we are after. It reveals that the computation of the energy release rate $-\partial \mathcal{W}^{{\rm Eq}}/\partial \mathrm{\Gamma}_0$ in the Griffith criticality condition (\ref{Gc-0}) for a trousers fracture test --- regardless of whether the test is carried out by prescribing a force $P(t)$ or a deformation $l(t)$ --- amounts to determining two quantities:
\begin{itemize}

\item{the stretch $\lambda(t)$ that results in the elastomer by subjecting it to uniaxial tension with the same force $P(t)$ that is prescribed or measured, if $l(t)$ is prescribed, in the test and}

\item{the equilibrium elastic force $P^{\rm{Eq}}$, as defined by (\ref{PEq}).}

\end{itemize}
While the stretch $\lambda(t)$ can be readily determined from a separate uniaxial tension test, the determination of $P^{\rm{Eq}}$ would appear to require, in principle, having access to the local deformation field in the trousers specimen at hand, which is only possible by solving in full the pertinent initial-boundary-value problem making use of an appropriate viscoelastic model for the elastomer. Fortunately, in practice, the equilibrium elastic force $P^{\rm{Eq}}$ can be determined directly in terms of a global measurement from the trousers test itself. This is because of two distinguishing properties of $P^{\rm{Eq}}$, which we outline next.

\begin{remark}\label{Remark1} $P^{\rm{Eq}}$ is the dominant term in (\ref{ERR-l}).  {\rm In practice, as already noticed by Rivlin and Thomas (1953) from their own experimental results, fracture in trousers specimens with sufficiently large height $H$ nucleates when the stretch in region $\texttt{B}$ is negligible, that is, when $\lambda(t)\approx 1$. This allows to simplify relation (\ref{ERR-l}) to
\begin{equation}
-\dfrac{\partial \mathcal{W}^{{\rm Eq}}}{\partial \mathrm{\Gamma}_0}=\dfrac{2}{B} P^{\rm{Eq}}.\label{ERR-2}
\end{equation}
}
\end{remark}

\begin{remark}\label{Remark2} $P^{\rm{Eq}}$ is substantially only a function of the global stretch $l(t)/l_0$ between the grips. {\rm For typical trousers specimens, for which relation (\ref{ERR-2}) applies, the full-field analysis presented in Subsection \ref{Sec:PEq} below reveals that  --- rather remarkably ---  the equilibrium elastic force $P^{\rm{Eq}}$ is \emph{de facto} only a function of the current value of the global stretch $l(t)/l_0$ between the grips, and hence, in particular,  independent of the length $A$ of the pre-existing crack and of the loading rate. With a slight abuse of notation, we write
\begin{equation}
P^{\rm{Eq}}=P^{\rm{Eq}}\left(\dfrac{l(t)}{l_0}\right). \label{PEq-lt}
\end{equation}
\emph{Ergo}, given an elastomer of interest, in order to determine $P^{\rm{Eq}}$ in practice, it would suffice to carry out a trousers test at a slow enough rate that viscous dissipation is negligible, measure both $P(t)$ and $l(t)/l_0$, and then use these measurements to determine the function (\ref{PEq-lt}), since  $P^{\rm{Eq}}=P(t)$ in the absence of viscous dissipation.

}
\end{remark}

\paragraph{A unique critical global stretch $l_c/l_0$} When combined with the Griffith criticality condition (\ref{Gc-0}), the facts that the energy release rate $-\partial \mathcal{W}^{{\rm Eq}}/\partial \mathrm{\Gamma}_0$ is given exclusively in terms of $P^{\rm{Eq}}$ and that $P^{\rm{Eq}}$ is only a function of the global stretch $l(t)/l_0$ imply that, for a given elastomer of interest, there is a unique critical global stretch
\begin{equation*}
\dfrac{l_c}{l_0}
\end{equation*}
at which fracture nucleates in a trousers test, irrespective of the length of the pre-existing crack and of the loading rate.

Interestingly, the above is the same type of unique criticality result that occurs in ``pure-shear'' fracture tests (Shrimali and Lopez-Pamies, 2023a).

Regrettably, virtually none of the experimental studies that have been reported in the literature for trousers fracture tests include measurements of the critical deformation $l_c$ at which fracture nucleation takes place. The sole exception that we are aware of is the original data of Rivlin and Thomas (1953), which pertains to natural-rubber specimens that contained pre-existing cracks of various different lengths $A$, but that were loaded at the same (slow) rate; see Figs. 2(ii) and 4(ii) in their work. Their results indicate (to within experimental error) that fracture nucleation occurs at a critical stretch $l_c/l_0$ that is indeed independent of the length of the pre-existing crack.

\subsection{Fracture propagation}\label{Sec: Propa}

The analysis presented in the preceding subsection pertains to the \emph{nucleation} of fracture, that is, the first instance at which new surface is created from the pre-existing crack. As first recognized by Greensmith and Thomas (1955), trousers fracture tests are also particularly useful to study fracture \emph{propagation} in viscoelastic elastomers. This is because, under the often satisfied twofold premise that
\begin{itemize}

\item{$\lambda(t)\approx 1$ and}

\item{crack propagation is a smooth process in time,}

\end{itemize}
the rate of crack propagation ${\rm d}\mathrm{\Gamma}(t)/{\rm d}t$ is given in terms of the rate ${\rm d}l(t)/{\rm d}t$ of separation between the grips by the simple relation
\begin{equation*}
\dfrac{{\rm d}\mathrm{\Gamma}}{{\rm d} t}(t)=\dfrac{B}{2}\dfrac{{\rm d} l}{{\rm d}t}(t),
\end{equation*}
or, equivalently, using the identity ${\rm d}\mathrm{\Gamma}(t)=B{\rm d}a(t)$, by the relation
\begin{equation}
\dfrac{{\rm d}a}{{\rm d} t}(t)=\dfrac{1}{2}\dfrac{{\rm d} l}{{\rm d}t}(t)\label{a-0}
\end{equation}
in terms of the current (undeformed) length $a(t)$ of the crack.

\paragraph{The experimentally prominent case of applied deformation $l(t)$ at a constant rate} For the case when the test is carried out by separating the grips at a constant rate $\dot{l}_0$, so that, again, $l(t)=l_0+\dot{l}_0 t$, it follows immediately from (\ref{a-0}) that
\begin{equation}
a(t)=A +\dfrac{\dot{l}_0}{2} t. \label{a-1}
\end{equation}

Granted the crack evolution (\ref{a-1}), note that
\begin{equation}
\dfrac{l(t)}{2 a(t)}=\dfrac{2A+\dot{l}_0 t}{2\left(A +\dfrac{\dot{l}_0}{2}t\right)}=1=\dfrac{l_c}{l_0} \label{lc-l0-a}
\end{equation}
for all $t>0$, where, we have made critical use of the inequality $B\ll A$.

The string of equalities (\ref{lc-l0-a}) reveals that the propagation (\ref{a-1}) of the crack in a trousers fracture test carried out by pulling the grips apart at a constant rate of deformation is such that the current global stretch $l(t)/2 a(t)$ between the grips is always at the critical global stretch $l_c/l_0$. In other words, the Griffith criticality condition  (\ref{Gc-0}) is constantly satisfied. It is for this reason that what is typically observed in experiments  --- whenever the crack propagation happens to be smooth in time --- is that the crack length evolves according to (\ref{a-1}) and that it does so at a constant force $P(t)$, of different value for different applied deformation rates $\dot{l}_0$; see, e.g., Greensmith and Thomas (1955), Mullins (1959), and Gent (1996).

\begin{remark}\label{Remark3} Lack of smoothness/continuity of crack propagation in time.  {\rm Already in their early pioneering experiments, Greensmith and Thomas (1955) noticed that crack propagation in trousers tests may not be smooth or even continuous in time. They referred to such a type of propagation as ``stick-slip''; see Fig. 3 in their work. For non-crystallizable elastomers --- in particular, for various types of SBR --- they observed that crack propagation is mostly smooth in time, except possibly at high rates of deformation and low temperatures. For natural rubber, the most prominent crystallizable elastomer, on the other hand, they observed that crack propagation is mostly discontinuous in time, irrespective of the loading rate and temperature.
}
\end{remark}

\section{Formulation of the initial-boundary-value problem for the trousers test}\label{Sec:Full-field}

Having analyzed the trousers fracture test from a global perspective, we now turn to its full-field analysis.

\subsection{Initial configuration}

Consider the rectangular specimens depicted in Fig. \ref{Fig1} of length $L=150$ mm and height $H=40$ mm in the $\bfe_3$ and $\bfe_1$ directions and constant thickness $B=1$ mm in the $\bfe_2$ direction. The specimens contain a pre-existing edge crack of five different lengths
\begin{equation*}
A=49,49.5,50,50.5,51\; {\rm mm}
\end{equation*}
in the $\bfe_3$ direction. These specific values for $L$, $H$, $B$, $A$ are chosen here because they are representative of those typically used in experiments; see, in particular, the classical experiments of Greensmith and Thomas (1955). Here, $\{\bfe_i\}$ stands for the laboratory frame of reference. We place its origin at the specimens' midplane along the bottom edge so that, in their $\texttt{i}$nitial configuration, the specimens occupy the domain
\begin{equation*}
\overline{\mathrm{\Omega}}^{\texttt{i}}_0=\{\bfX: \bfX\in\mathcal{P}^{\texttt{i}}_0\setminus\mathrm{\Gamma}^{\texttt{i}}_0\},
\end{equation*}
where
\begin{equation*}
\mathcal{P}^{\texttt{i}}_0=\left\{\bfX: |X_1|\leq\dfrac{H}{2},\,|X_2|\leq\dfrac{B}{2},\,   -L\leq X_3 \leq 0  \right\}
\end{equation*}
and
\begin{equation*}
\mathrm{\Gamma}^{\texttt{i}}_0=\left\{\bfX: X_1=0,\,|X_2|\leq\dfrac{B}{2},\,  -L\leq X_3 \leq A-L  \right\}.
\end{equation*}

\subsection{Reference configuration and kinematics}\label{Sec: Reference Config}

As already noted in the preceding global analysis of the problem, in an actual trousers fracture test, the specimen is mounted in the testing machine by bending in opposite directions the two legs of the trousers until they are brought to lie in the same plane, at which point their bottom ends are firmly gripped and the specimen is ready to be loaded. For this reason, it proves convenient not to use the initial configuration as the reference configuration to carry out the analysis, but to use, instead, the configuration as initially mounted in the testing machine.

\begin{remark}\label{Remark4} Residual stresses in the reference configuration.  {\rm In the reference configuration identified above, there are residual stresses due to the bending of the legs of the trousers. Numerical simulations show that these have no significant impact on the response of the specimen when the grips are pulled apart and hence that they can be neglected altogether; this is hardly surprising, since the specimens are very thin ($H/B=40$).
}
\end{remark}

In our analysis, granted that residual stresses can be neglected, we therefore take the reference configuration to be both undeformed and stress free. Specifically, at time $t=0$, in their reference configuration, we consider that the specimens occupy the domain
\begin{equation*}
\overline{\mathrm{\Omega}}_0=\overline{\mathrm{\Omega}}^{\mathcal{L}}_0\cup\overline{\mathrm{\Omega}}^{\mathcal{R}}_0\cup\overline{\mathrm{\Omega}}^{\mathcal{B}}_0,
\end{equation*}
with $\overline{\mathrm{\Omega}}^{\mathcal{L}}_0=\overline{\mathrm{\Omega}}^{\mathcal{L}_s}_0\cup\overline{\mathrm{\Omega}}^{\mathcal{L}_f}_0$ and $\overline{\mathrm{\Omega}}^{\mathcal{R}}_0=\overline{\mathrm{\Omega}}^{\mathcal{R}_s}_0\cup\overline{\mathrm{\Omega}}^{\mathcal{R}_f}_0$, where
\begin{align*}
\Omega^{\mathcal{L}_s}_0=&\left\{\bfX: -\frac{H}{2}\leq X_1< 0,\,-A-\frac{3B}{2}\leq X_2\leq -\frac{3B}{2},\right.\\
& \hspace{0.4cm}\left.A-L-2B\leq X_3\leq A-L-B  \right\},\\
\Omega^{\mathcal{L}_f}_0=&\left\{\bfX: -\frac{H}{2}\leq X_1< 0,-\frac{3B}{2}\leq\,X_2,\,X_3\leq A-L,\right.\\
& \hspace{0.4cm}\left.B\leq\sqrt{\left(X_2+\frac{3B}{2}\right)^2+(X_3+L-A)^2}\leq 2B \right\},\\
\Omega^{\mathcal{R}_s}_0=&\left\{\bfX: 0< X_1\leq \frac{H}{2},\,B\leq X_2\leq A+B,\right.\\
& \hspace{0.4cm}\left.A-L-2B\leq X_3\leq A-L-B  \right\},\\
\Omega^{\mathcal{R}_f}_0=&\left\{\bfX:0< X_1\leq \frac{H}{2},\, X_2\leq \frac{3B}{2},\,X_3\leq A-L,\right.\\
& \hspace{0.4cm}\left.B\leq\sqrt{\left(X_2-\frac{3B}{2}\right)^2+(X_3+L-A)^2}\leq 2B \right\},\\
\Omega^{\mathcal{B}}_0=&\left\{\bfX: -\frac{H}{2}\leq X_1\leq \frac{H}{2},\,-\frac{B}{2}\leq X_2\leq \frac{B}{2},\right.\\
& \hspace{0.4cm}\left.A-L\leq X_3\leq 0  \right\}.
\end{align*}

\begin{remark}\label{Remark5} The fillet in the reference configuration.  {\rm In the reference configuration defined by the above domains, the pre-existing crack has been assumed to feature a circular fillet of inner radius $B=1$ mm and so its initial inner length is \emph{not} simply $A$ but $A+\pi B$. Numerical simulations show that the specifics of the fillet have no significant impact on the response of the specimens. Again, this is because $B\ll H<A$. For notational simplicity, in the sequel, we will continue referring to $A$ as the initial length of the pre-existing crack with the understanding that its actual initial length contains a correction of order $B$.
}
\end{remark}

At a later time $t\in(0,T]$, in response to the applied boundary conditions described below, the position vector $\bfX$ of a material point in the specimens will move to a new position specified by
\begin{equation*}
\bfx=\bfy(\bfX, t),
\end{equation*}
where $\bfy$ is a mapping from $\mathrm{\Omega}_0$ to the current configuration $\mathrm{\Omega}(t)$. We consider only invertible deformations, and write the deformation gradient field at $\bfX$ and $t$ as
\begin{equation*}
\bfF(\bfX, t)=\nabla\bfy(\bfX,t)=\frac{\partial \bfy}{\partial \bfX}(\bfX,t).
\end{equation*}

\subsection{Constitutive behavior of the elastomer}

The specimens, again, are taken to be made of an isotropic incompressible elastomer. So as to uncover the key features of the trousers fracture test within the simplest of settings, the viscoelastic behavior of this elastomer is taken to be canonical in the sense that its elasticity is Gaussian and its viscosity is constant. Precisely, making use of the formulation of Kumar and Lopez-Pamies (2016), the first Piola-Kirchhoff stress tensor $\bfS$ at any material point $\bfX\in\mathrm{\Omega}_0$ and time $t\in[0,T]$ is given by the relation
\begin{equation}
\bfS(\bfX,t)=\mu\bfF+\nu\bfF{\bfC^v}^{-1}-p\bfF^{-T},\label{S-KLP}
\end{equation}
where the internal variable $\bfC^v$ is defined implicitly as the solution of the evolution equation
\begin{equation}
\dot{\bfC}^v(\bfX,t)=\dfrac{\nu}{\eta}\left[\bfC-\dfrac{1}{3}\left(\bfC\cdot{\bfC^v}^{-1}\right)\bfC^v\right],  \label{Evolution-KLP}
\end{equation}
$p$ stands for the arbitrary hydrostatic pressure associated with the incompressibility constraint $J=\det\bfF=1$, $\bfC=\bfF^T\bfF$ denotes the right Cauchy-Green deformation tensor, the ``dot'' notation stands for the Lagrangian time derivative (i.e., with $\bfX$ held fixed), and $\mu\geq 0$, $\nu\geq 0$, and $\eta\geq 0$ are three material constants. Specifically, $\mu$ denotes the initial shear modulus associated with the Gaussian elasticity of the elastomer at states of thermodynamic equilibrium, $\nu$ denotes the initial shear modulus associated with its additional Gaussian elasticity at non-equilibrium states, while $\eta$ stands for its viscosity.

\begin{figure}[t!]\centering
 \includegraphics[width=2.4in]{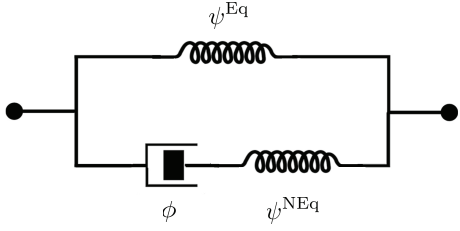}
\caption{\small The rheological representation of the viscoelastic model (\ref{S-KLP})-(\ref{Evolution-KLP}).}
   \label{Fig2}
\end{figure}

For later reference, beyond the preceding brief description, it is also appropriate to recall that the constitutive model (\ref{S-KLP})-(\ref{Evolution-KLP}) corresponds to a generalization of the classical Zener or standard solid model (Zener, 1948) to the setting of  finite deformations. Specifically, as schematically depicted by its rheological representation in Fig. \ref{Fig2}, the viscoelastic model (\ref{S-KLP})-(\ref{Evolution-KLP}) describes a solid that stores equilibrium and non-equilibrium elastic energy according to the free energy
\begin{equation}
\psi=\left\{\begin{array}{ll}\underbrace{\dfrac{\mu}{2}\left[I_1-3\right]}_\text{$\psi^{{\rm Eq}}(I_1)$}+\underbrace{\dfrac{\nu}{2}\left[I^e_1-3\right]}_\text{$\psi^{{\rm NEq}}\left(I^e_1\right)$}& {\rm if}\quad J=1\\ \\
+\infty & {\rm otherwise}\end{array}\right.,\label{free-energy}
\end{equation}
where $I_1={\rm tr}\,\bfC$ and $I_1^e={\rm tr}(\bfC{\bfC^{v}}^{-1})$, and dissipates energy according to the dissipation potential
\begin{equation*}
\phi=\left\{\hspace{-0.15cm}\begin{array}{ll}
\dfrac{\eta}{4}\,{\rm tr}\left(\dot{\bfC}^v{\bfC^v}^{-1}\dot{\bfC}^v{\bfC^v}^{-1}\right) & {\rm if}\;\det\bfC^v=1 \vspace{0.2cm} \\
+\infty &  {\rm otherwise}\end{array}\right. .
\end{equation*}

\subsection{Initial and boundary conditions}

As assumed from the outset in Subsection \ref{Sec: Reference Config}, the reference configuration is undeformed and stress free. Therefore, we have the initial conditions
\begin{equation}
\left\{\begin{array}{l}
\bfy(\bfX,0)=\bfX\vspace{0.2cm}\\
p(\bfX,0)=\mu+\nu\vspace{0.2cm}\\
\bfC^v(\bfX,0)=\bfI \end{array}\right., \quad\bfX\in \overline{\mathrm{\Omega}}_0\label{ICs}
\end{equation}
for the deformation field $\bfy(\bfX,t)$, the pressure field $p(\bfX,t)$, and the internal variable $\bfC^v(\bfX,t)$.

Save for the left grip boundary
\begin{align*}
\partial\Omega^{\mathcal{L}}_0=&\left\{\bfX: -\dfrac{H}{2}<X_1< 0,\,X_2=-A-\frac{3B}{2},\right.\\
& \hspace{0.7cm}\left.A-L-2B\leq X_3\leq A-L-B   \right\}
\end{align*}
and the right grip boundary
\begin{align*}
\partial\Omega^{\mathcal{R}}_0=&\left\{\bfX:  0<X_1\leq \dfrac{H}{2},\,X_2=A+\frac{3B}{2},\right.\\
& \hspace{0.7cm}\left.  A-L-2B\leq X_3\leq A-L-B   \right\},
\end{align*}
the entire boundary $\partial\Omega_0$ of the specimens is traction free. The left and right grip boundaries are separated in the $\bfe_2$ direction at the constant rate $\dot{l}_0$ so that, as a function of time $t\in[0,T]$, the current separation between the grips is given by the relation $l(t)=l_0+\dot{l}_0 t$, where $l_0=2A+3B$. Precisely, making use of the notation $\textbf{s}(\bfX,t)=\bfS\bfN$, we have that the boundary conditions in full read

\begin{equation}
\left\{\hspace{-0.15cm}\begin{array}{ll}
y_1(\bfX,t)=X_1, & (\bfX,t)\in\partial\Omega^{\mathcal{L}}_0\times[0,T] \vspace{0.15cm}\\
y_2(\bfX,t)=X_2-\dfrac{\dot{l}_0}{2}t, & (\bfX,t)\in\partial\Omega^{\mathcal{L}}_0\times[0,T] \vspace{0.15cm}\\
y_3(\bfX,t)=X_3, & (\bfX,t)\in\partial\Omega^{\mathcal{L}}_0\times[0,T] \vspace{0.15cm}\\
y_1(\bfX,t)=X_1, & (\bfX,t)\in\partial\Omega^{\mathcal{R}}_0\times[0,T] \vspace{0.15cm}\\
y_2(\bfX,t)=X_2+\dfrac{\dot{l}_0}{2}t, & (\bfX,t)\in\partial\Omega^{\mathcal{R}}_0\times[0,T] \vspace{0.15cm}\\
y_3(\bfX,t)=X_3, & (\bfX,t)\in\partial\Omega^{\mathcal{R}}_0\times[0,T] \vspace{0.15cm}\\
\textbf{s}=\textbf{0}, & \hspace{-1.75cm}(\bfX,t)\in\partial\Omega_0\setminus\left(\partial\Omega^{\mathcal{L}}_0\cup\partial\Omega^{\mathcal{R}}_0\right)\times[0,T]
\end{array}\right. ,\label{BCs}
\end{equation}
where $\bfN$ stands for the outward unit normal to the boundary $\partial\Omega_0$.

\subsection{Governing equations}

At this stage, we are in a position to put all the above ingredients together into a complete set of governing equations that describes the mechanical response of the specimens. In the absence of inertia and body forces, the resulting governing equations are nothing more than the equilibrium and incompressibility constraint equations
\begin{equation}
\left\{\begin{array}{ll}{\rm Div}\,\bfS={\bf0}, & \quad (\bfX,t)\in\mathrm{\Omega}_0\times[0,T] \vspace{0.2cm} \\
\det\nabla\bfy=1, & \quad (\bfX,t)\in\mathrm{\Omega}_0\times[0,T]
\end{array}\right. \label{Equilibrium-PDE}
\end{equation}
subject to the initial and boundary conditions (\ref{ICs})$_{1,2}$ and (\ref{BCs}), where $\bfS(\bfX,t)=\mu\nabla\bfy+\nu\nabla\bfy{\bfC^v}^{-1}-p\nabla\bfy^{-T}$, coupled with the evolution equation
\begin{align}
\dot{\bfC}^v=\dfrac{\nu}{\eta}\left[\nabla\bfy^T\nabla\bfy-\dfrac{1}{3}\left(\nabla\bfy^T\nabla\bfy\cdot{\bfC^v}^{-1}\right)\bfC^v\right], \label{Evolution-ODE}
\end{align}
subject to the initial condition (\ref{ICs})$_3$, for the deformation field $\bfy(\bfX,t)$, the pressure field $p(\bfX,t)$, and the internal variable $\bfC^v(\bfX,t)$.

In general, the initial-boundary-value problem (\ref{Equilibrium-PDE})-(\ref{Evolution-ODE}) with (\ref{ICs})-(\ref{BCs}) does not admit analytical solutions and hence must be solved numerically. All the results that we present below are generated by a variant of the numerical scheme introduced by Ghosh et al. (2021), which is based on a Crouzeix-Raviart finite-element discretization of space and a high-order explicit Runge-Kutta discretization of time.

\section{Results for a canonical elastomer with Gaussian elasticity and constant viscosity}\label{Results Gaussian}

In this section, we present solutions for the initial-boundary-value problem (\ref{Equilibrium-PDE})-(\ref{Evolution-ODE}) with (\ref{ICs})-(\ref{BCs}) describing the trousers fracture test of a canonical elastomer with equilibrium initial shear modulus
\begin{equation*}
\mu=1\;{\rm MPa},
\end{equation*}
three different non-equilibrium initial shear moduli
\begin{equation*}
\nu=2, 5, 10\; {\rm MPa},
\end{equation*}
and three different viscosities
\begin{equation*}
\eta=5,25,100\; {\rm MPa}\,{\rm s}.
\end{equation*}
These ranges of values are chosen here because they are prototypical of standard elastomers. Note, in particular, that they correspond to elastomers with relaxation times $\tau=\eta/\nu=0.5$, $1$, $2.5$, $5$, $10$, $12.5$, $20$, $50$ ${\rm s}$.

The solutions pertain to deformation rates in the range
\begin{equation*}
\dot{l}_0\in[2\times 10^{-4}, 100]\; {\rm mm}/{\rm s}
\end{equation*}
spanning more than five orders of magnitude. This ensures that the entire spectrum of behaviors --- from elasticity-dominated to viscosity-dominated --- is probed.

\subsection{The force-deformation response}\label{Sec:force}

\begin{figure}[b!]
  \subfigure[]{
   \label{fig:3a}
   \begin{minipage}[]{0.5\textwidth}
   \centering \includegraphics[width=2.3in]{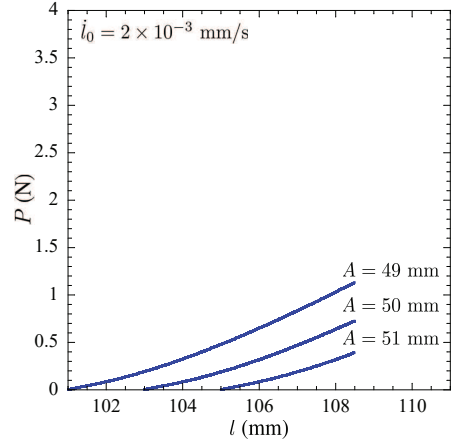}
   \vspace{0.2cm}
   \end{minipage}}
  \subfigure[]{
   \label{fig:3b}
   \begin{minipage}[]{0.5\textwidth}
   \centering \includegraphics[width=2.3in]{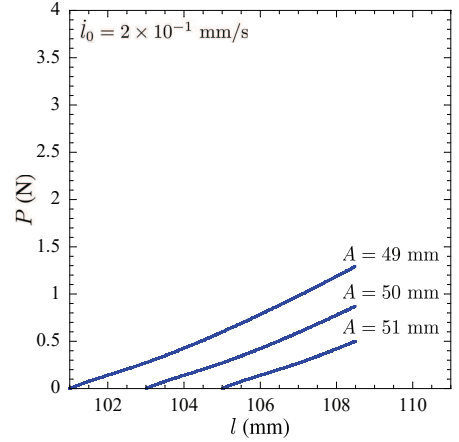}
   \vspace{0.2cm}
   \end{minipage}}
     \subfigure[]{
   \label{fig:3b}
   \begin{minipage}[]{0.5\textwidth}
   \centering \includegraphics[width=2.3in]{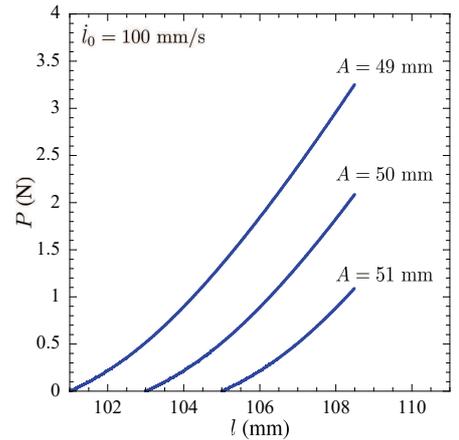}
   \vspace{0.2cm}
   \end{minipage}}
   \caption{Force-deformation response of trousers specimens with non-equilibrium shear modulus $\nu=2$ MPa, viscosity $\eta=5$ ${\rm MPa}\, {\rm s}$, and pre-existing cracks of various lengths $A$. Parts (a), (b), and (c) show results for deformations applied at the constant rates $\dot{l}_0=2\times10^{-3}$ mm/s$^{-1}$, $\dot{l}_0=2\times10^{-1}$ mm/s$^{-1}$, and $\dot{l}_0=100$ mm/s$^{-1}$, respectively.}\label{Fig3}
\end{figure}

Figure \ref{Fig3} presents results for the total force $P(t)$ required to deform the specimens with non-equilibrium shear modulus $\nu=2$ MPa, viscosity $\eta=5$ ${\rm MPa}\, {\rm s}$, and pre-existing cracks of length $A=49, 50, 51$ mm at three constant deformation rates $\dot{l}_0$. The results are shown for $P(t)$ as a function of the applied deformation $l(t)$ for $\dot{l}_0=2\times10^{-3}$ mm/s$^{-1}$ in part (a), $\dot{l}_0=2\times10^{-1}$ mm/s$^{-1}$ in part (b), and $\dot{l}_0=100$ mm/s$^{-1}$ in part (c).

There are three observations worth pointing out. First, specimens with larger cracks require smaller forces to reach the same deformation. Second, larger forces are required to reach a given deformation applied at a higher deformation rate. Finally, all force-deformation responses exhibit some nonlinearity at the beginning of the loading process, whereas they are mostly linear at larger deformations.

%
%

\subsection{The total deformation energy $\mathcal{W}$ and its partition into $\mathcal{W}^{{\rm Eq}}$, $\mathcal{W}^{{\rm NEq}}$, and $\mathcal{W}^{v}$}\label{Sec:W}

%
\begin{figure}[t!]
  \subfigure[]{
   \label{fig:4a}
   \begin{minipage}[]{0.5\textwidth}
   \centering \includegraphics[width=3in]{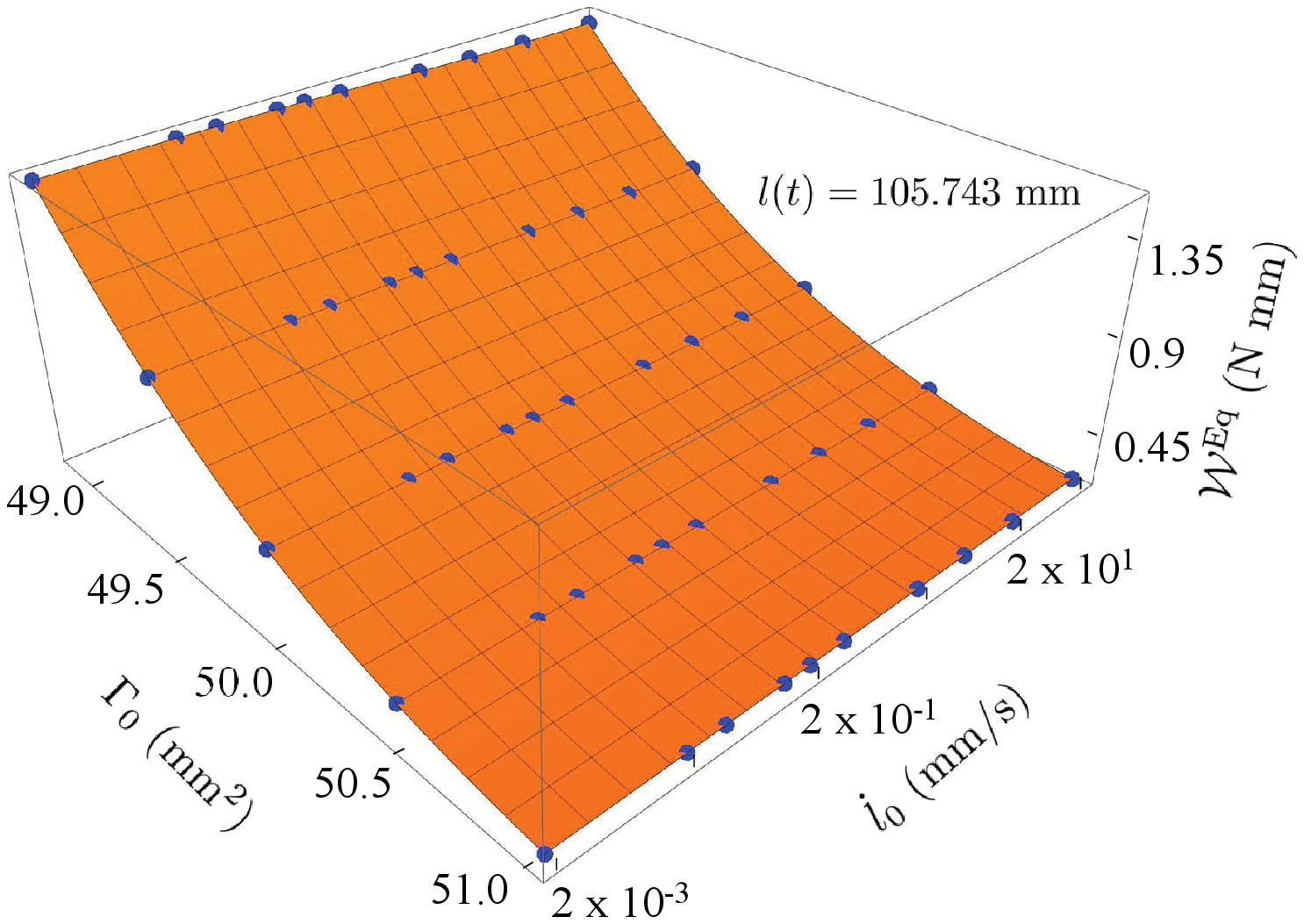}
   \vspace{0.2cm}
   \end{minipage}}
  \subfigure[]{
   \label{fig:4b}
   \begin{minipage}[]{0.5\textwidth}
   \centering \includegraphics[width=3in]{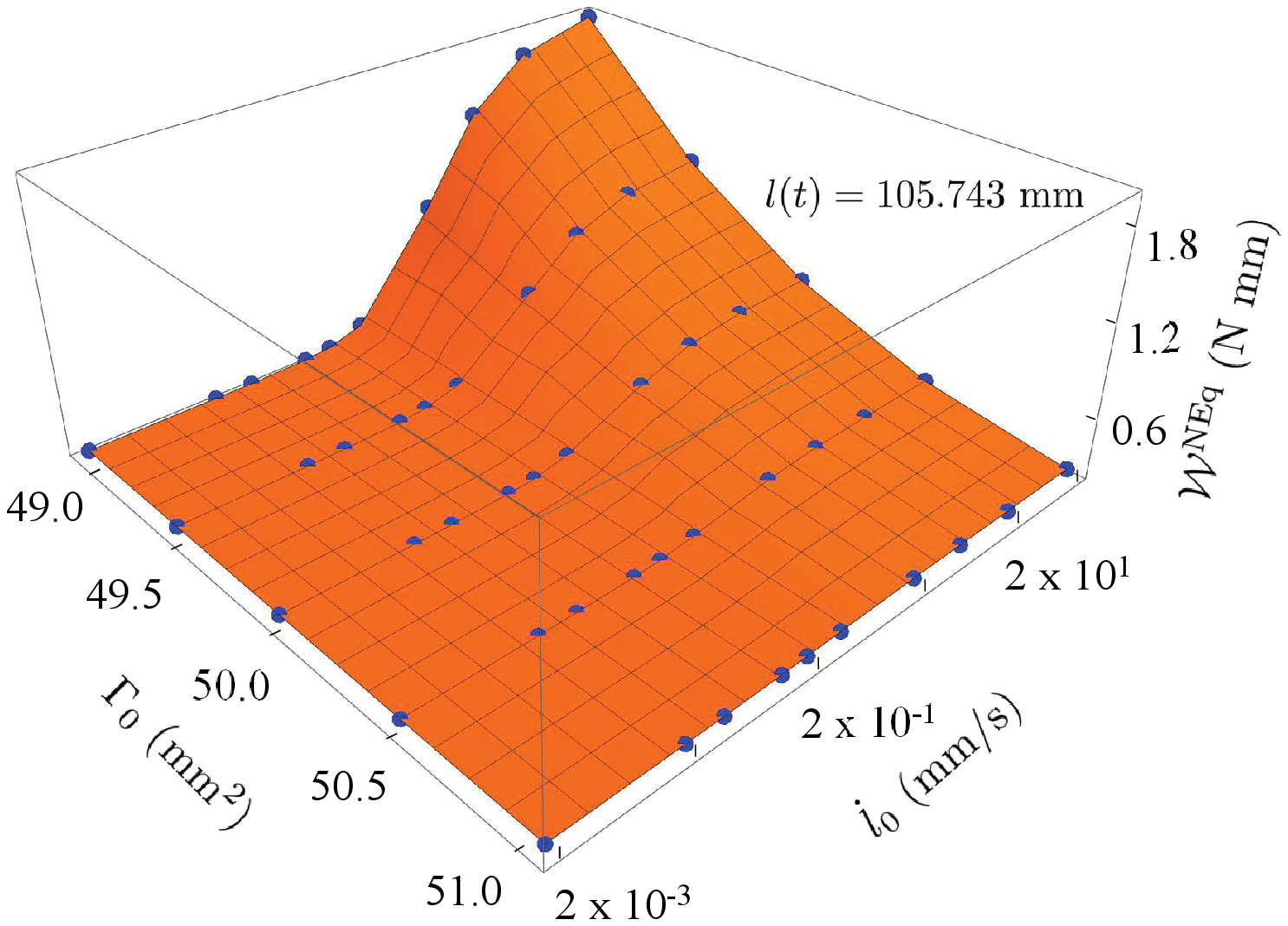}
   \vspace{0.2cm}
   \end{minipage}}
     \subfigure[]{
   \label{fig:4c}
   \begin{minipage}[]{0.5\textwidth}
   \centering \includegraphics[width=3in]{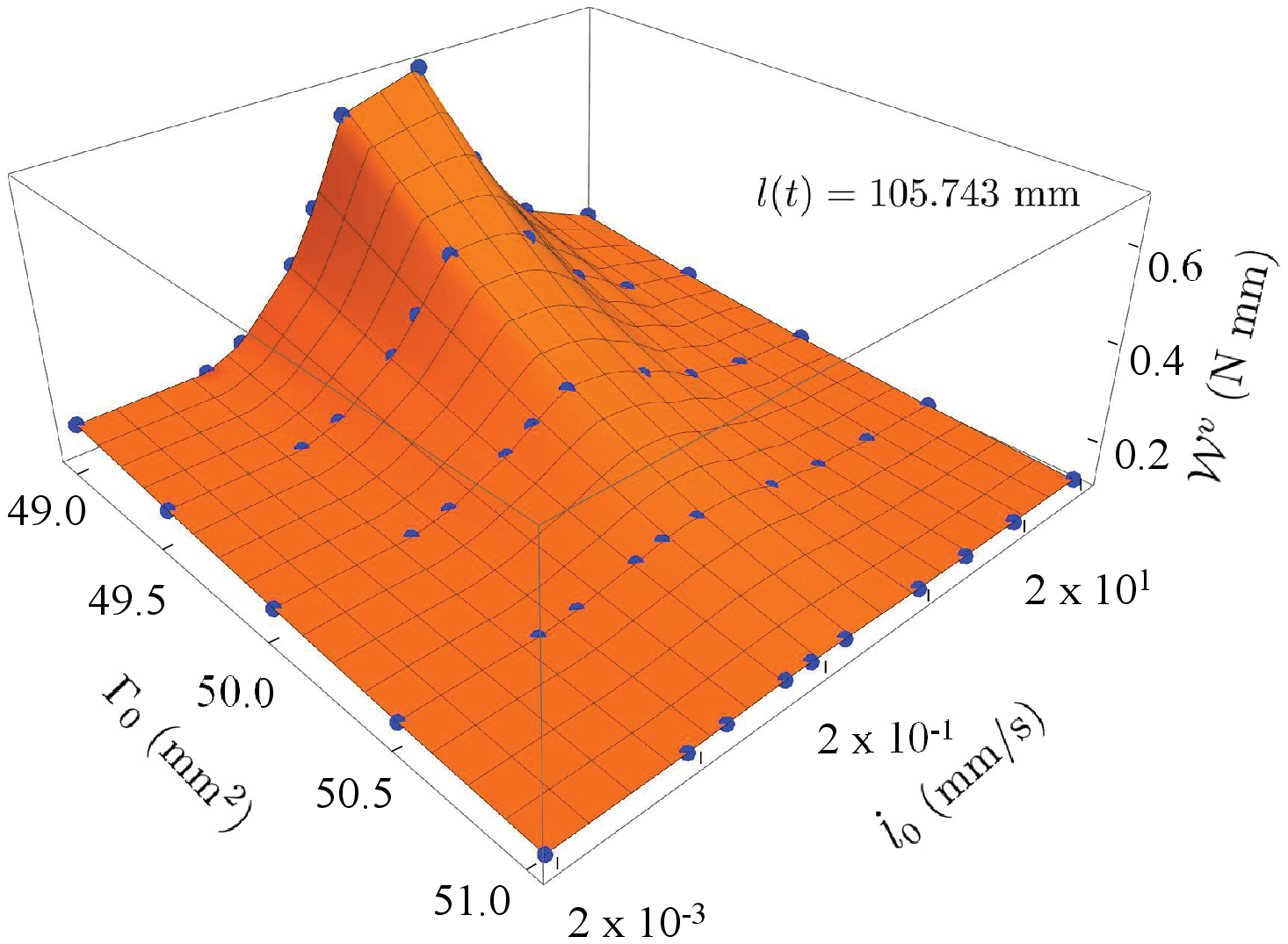}
   \vspace{0.2cm}
   \end{minipage}}
   \caption{Computed values from (\ref{WEq-NH})-(\ref{Wv-NH}) of (a) the equilibrium elastic energy $\mathcal{W}^{{\rm Eq}}$, (b) the non-equilibrium elastic energy $\mathcal{W}^{{\rm NEq}}$, and (c) the dissipated viscous energy $\mathcal{W}^{v}$ in trousers specimens with non-equilibrium shear modulus $\nu=2$ MPa, viscosity $\eta=5$ ${\rm MPa}\, {\rm s}$, deformed at $l(t)=105.743$ mm, plotted as functions of the initial crack surface $\Gamma_0=A \times B$ and the applied deformation rate $\dot{l}_0$.}\label{Fig4}
\end{figure}
%

The areas under the curves in the results presented in Fig. \ref{Fig3} correspond to the total work done externally by the grips and, consequently, they correspond as well to the total deformation energy stored and dissipated by the elastomer. We thus have
\begin{equation*}
\mathcal{W}=\displaystyle\int_{l_0}^{l_0+\dot{l}_0 t} P\,{\rm d}l.
\end{equation*}
Given that the elastomer is a Gaussian elastomer with constant viscosity, we also have that
\begin{align}
\mathcal{W}^{{\rm Eq}}=\displaystyle\int_{\Omega_0}\psi^{{\rm Eq}}(I_1)\,{\rm d}\bfX=\displaystyle\int_{\Omega_0}\dfrac{\mu}{2}\left[{\rm tr}\,\bfC-3\right]\,{\rm d}\bfX,\label{WEq-NH}
\end{align}
\begin{align}
\mathcal{W}^{{\rm NEq}}=\displaystyle\int_{\Omega_0}\psi^{{\rm NEq}}(I_1^e)\,{\rm d}\bfX=\displaystyle\int_{\Omega_0}\dfrac{\nu}{2}\left[{\rm tr}(\bfC{\bfC^{v}}^{-1})-3\right]\,{\rm d}\bfX, \label{WNEq-NH}
\end{align}
and
\begin{align}
\mathcal{W}^{v}=&\mathcal{W}-\mathcal{W}^{{\rm Eq}}-\mathcal{W}^{{\rm NEq}}, \label{Wv-NH}
\end{align}
in terms of the equilibrium and non-equilibrium parts of the free energy (\ref{free-energy}).

Figure \ref{Fig4} shows results for $\mathcal{W}^{{\rm Eq}}$, $\mathcal{W}^{{\rm NEq}}$, and $\mathcal{W}^{v}$ --- as computed from expressions (\ref{WEq-NH})-(\ref{Wv-NH}) and the pertinent numerical solutions for the deformation field $\bfy(\bfX,t)$ and internal variable $\bfC^v(\bfX,t)$ --- at the deformation $l(t)=105.743$ mm, plotted as functions of the initial crack surface $\Gamma_0=A \times B$ and the deformation rate $\dot{l}_0$. The results at other fixed values of the deformation $l(t)$ are not fundamentally different from those shown in Fig. \ref{Fig4} for $l(t)=105.743$ mm, which therefore can be viewed as representative of those at any deformation $l(t)$.

The following comments are in order. With the sole exception of the dependence of the equilibrium elastic energy $\mathcal{W}^{{\rm Eq}}$ on the rate $\dot{l}_0$ of the applied deformation, all three parts of the deformation energy appear to depend nonlinearly on both the crack surface $\mathrm{\Gamma}_0$ and $\dot{l}_0$. Distinctly, with respect to $\dot{l}_0$, both the non-equilibrium energy $\mathcal{W}^{{\rm NEq}}$ and the viscous dissipated energy $\mathcal{W}^{v}$  appear to be bounded, but whereas $\mathcal{W}^{{\rm NEq}}$ increases monotonically with increasing $\dot{l}_0$, $\mathcal{W}^{v}$ exhibits a $\Gamma_0$-dependent local maximum away from which $\mathcal{W}^{v}$ becomes vanishingly small.

\subsection{The derivative $-\partial\mathcal{W}^{{\rm Eq}}/\partial\Gamma_0$}\label{Sec:PEq}

%
\begin{figure}[b!]
   \centering \includegraphics[width=2.4in]{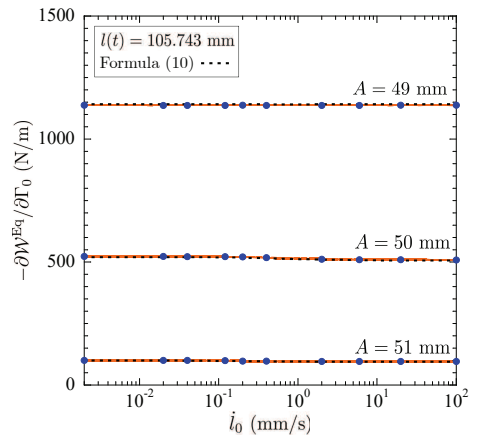}
   \vspace{0.2cm}
   \caption{Computed values from Fig. \ref{Fig4}(a) of the energy release rate $-\partial\mathcal{W}^{{\rm Eq}}/\partial\Gamma_0$ for trousers specimens with non-equilibrium shear modulus $\nu=2$ MPa, viscosity $\eta=5$ ${\rm MPa}\, {\rm s}$, and pre-existing cracks of length $A=49, 50, 51$ mm deformed at $l(t)=105.743$ mm, plotted as functions of the applied deformation rate $\dot{l}_0$. For direct comparison, the results obtained with the approximate formula (\ref{ERR-2}) are also plotted (dotted lines). }\label{Fig5}
\end{figure}
%

From the type of the 3D plot presented in Fig. \ref{Fig4}(a), we can compute numerically the energy release rate $-\partial\mathcal{W}^{{\rm Eq}}/\partial\Gamma_0$ entering the Griffith criticality condition (\ref{Gc-0}). Figure \ref{Fig5} reports such a computation of $-\partial\mathcal{W}^{{\rm Eq}}/\partial\Gamma_0$ in terms of the applied deformation rate $\dot{l}_0$ for specimens with non-equilibrium shear modulus $\nu=2$ MPa, viscosity $\eta=5$ ${\rm MPa}\, {\rm s}$, and pre-existing cracks of length $A=49, 50, 51$ mm at the same fixed deformation $l(t)=105.743$ mm considered in Fig. \ref{Fig4}(a). For direct comparison, the results produced by the approximate formula (\ref{ERR-2}) are also included in the figure (dotted lines).

We remark that, much like the results presented in Fig. \ref{Fig4}(a) are representative of any fixed value of $l(t)$, the results for $-\partial\mathcal{W}^{{\rm Eq}}/\partial\Gamma_0$ at other fixed values of the deformation $l(t)$ are qualitatively the same as those shown in Fig. \ref{Fig5} for $l(t)=105.743$ mm.

%
\begin{figure}[t!]
  \subfigure[]{
   \label{fig:5a}
   \begin{minipage}[]{0.5\textwidth}
   \centering \includegraphics[width=2.4in]{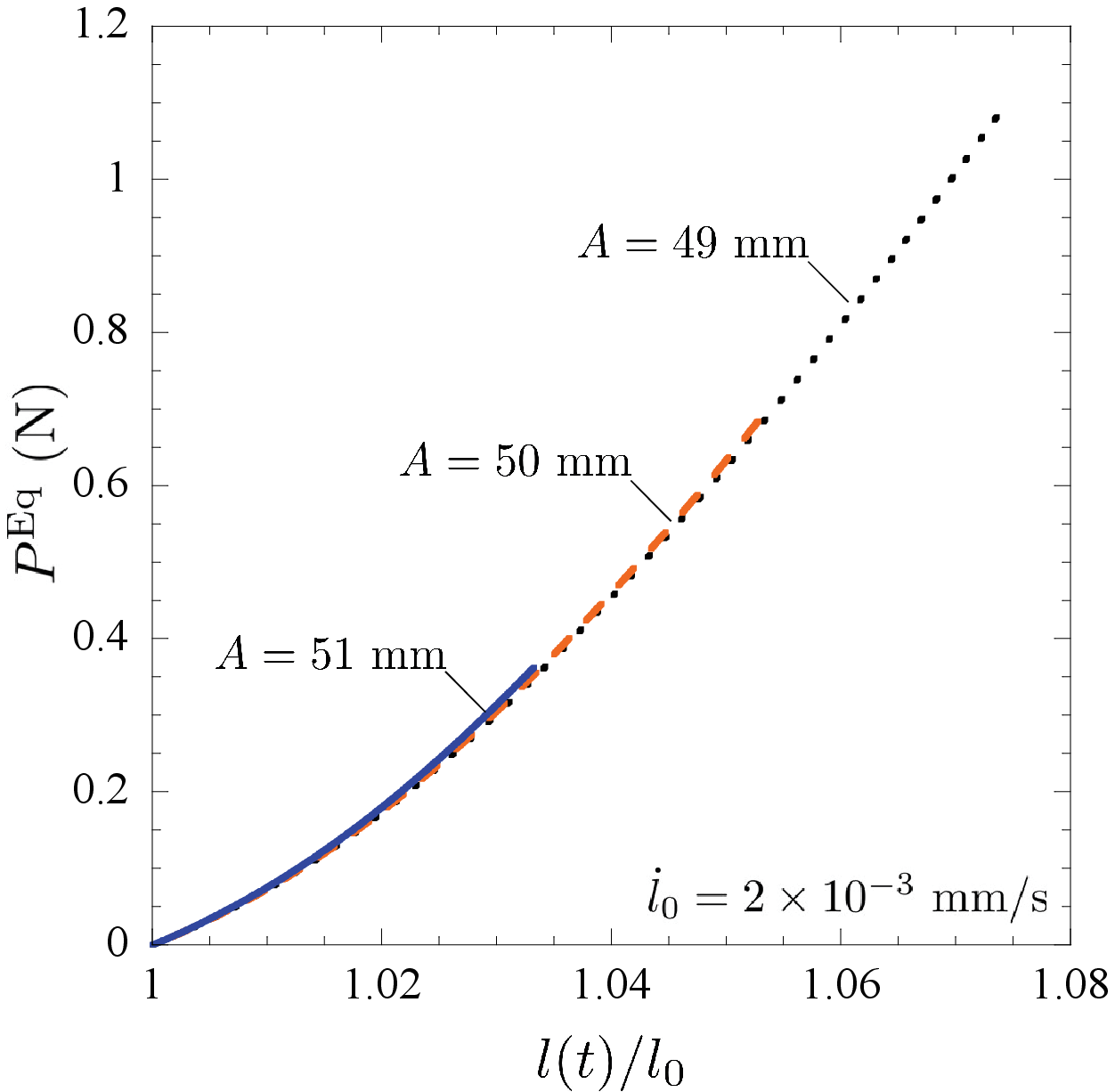}
   \vspace{0.2cm}
   \end{minipage}}
  \subfigure[]{
   \label{fig:5b}
   \begin{minipage}[]{0.5\textwidth}
   \centering \includegraphics[width=2.4in]{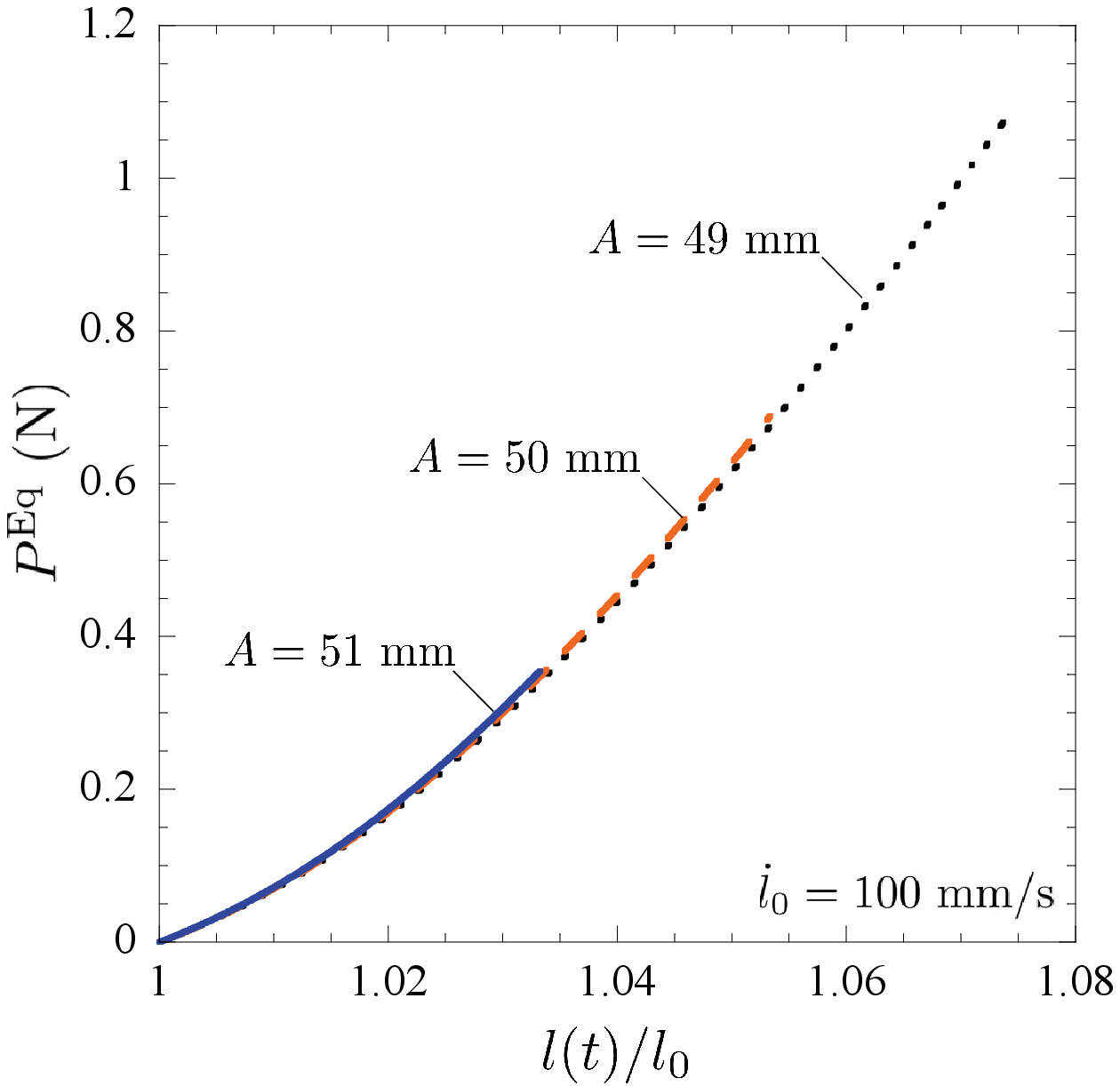}
   \vspace{0.2cm}
   \end{minipage}}
   \caption{The equilibrium elastic force (\ref{PEq}) as a function of the global stretch $l(t)/l_0$ between the grips for trousers specimens with non-equilibrium shear modulus $\nu=2$ MPa, viscosity $\eta=5$ ${\rm MPa}\, {\rm s}$, and pre-existing cracks of various lengths $A$. Parts (a) and (b) show results for deformations applied at the constant rates $\dot{l}_0=2\times10^{-3}$ mm/s$^{-1}$ and $\dot{l}_0=100$ mm/s$^{-1}$, respectively.}\label{Fig6}
\end{figure}
%

A key observation from Fig. \ref{Fig5} is that the energy release rate $-\partial\mathcal{W}^{{\rm Eq}}/\partial\Gamma_0$ is essentially independent of the rate $\dot{l}_0$ at which the grips are separated. On the other hand, $-\partial\mathcal{W}^{{\rm Eq}}/\partial\Gamma_0$ does depend on the length $A$ of the pre-existing crack, in particular, it increases with decreasing $A$. Another key observation from Fig. \ref{Fig5} is that the formula (\ref{ERR-2}) provides a good approximation for $-\partial\mathcal{W}^{{\rm Eq}}/\partial\Gamma_0$.

To gain further insight into the energy release rate $-\partial\mathcal{W}^{{\rm Eq}}/\partial\Gamma_0$ and its approximate representation (\ref{ERR-2}), Fig. \ref{Fig6} presents results for the equilibrium elastic force $P^{{\rm Eq}}$ as a function of the global stretch $l(t)/l_0$ between the grips for trousers specimens with non-equilibrium shear modulus $\nu=2$ MPa, viscosity $\eta=5$ ${\rm MPa}\, {\rm s}$, and pre-existing cracks of length $A=49, 50, 51$ mm. The results are shown for  $\dot{l}_0=2\times10^{-3}$ mm/s$^{-1}$ in part (a) and $\dot{l}_0=100$ mm/s$^{-1}$ in part (b), that is, the slowest and fastest deformation rates $\dot{l}_0$ considered in Fig. $\ref{Fig5}$.

Foremost, as announced in Remark \ref{Remark2}, the results in Fig. \ref{Fig6} show that the equilibrium elastic force $P^{{\rm Eq}}$ --- and hence, as per formula (\ref{ERR-2}), the energy release rate $-\partial\mathcal{W}^{{\rm Eq}}/\partial\Gamma_0$ --- is \emph{de facto} only a function of the global stretch $l(t)/l_0$ and hence, in particular, independent of the length $A$ of the pre-existing crack as well as of the rate $\dot{l}_0$ at which the deformation is applied.

\paragraph{A unique critical global stretch $l_c/l_0$} When combined with the Griffith criticality condition (\ref{Gc-0}) --- consistent with the conclusion established from the global analysis in Section \ref{Sec:Global} --- the full-field results presented in Figs. \ref{Fig5} and \ref{Fig6} for a canonical elastomer imply that, indeed, there is a unique critical global stretch
$$\dfrac{l_c}{l_0}$$
at which fracture nucleates in a trousers test and that this critical global stretch is independent of the length of the pre-existing crack and of the deformation rate.

To see this via an example, consider that the viscoelastic behavior of the elastomer being tested can be described by the canonical behavior assumed in this section, with equilibrium shear modulus $\mu=1$ MPa, non-equilibrium shear modulus $\nu=2$ MPa, and viscosity $\eta=5$ ${\rm MPa}\, {\rm s}$. Consistent with the standard range (\ref{Gc-100}) for common elastomers, consider further that the intrinsic fracture energy of the elastomer being tested is $G_c=100$ N/m. Then, according to the Griffith criticality condition (\ref{Gc-0}) and the formula (\ref{ERR-2}), fracture nucleation will occur whenever $P^{{\rm Eq}}=B G_c/2=(10^{-3}\,{\rm m})\times (100\, {\rm N}/{\rm m})/2=0.05$ N. In turn, according to the results in Fig. \ref{Fig6}, fracture nucleation will occur at the critical global stretch $l_c/l_0=1.0071$, irrespective of the length $A$ of the pre-existing crack and of the deformation rate $\dot{l}_0$ used to carry out the test.

\subsection{The critical tearing energy $T_c$}\label{Sec:Tc-Canon}

Following in the footstep of Greensmith and Thomas (1955), the vast majority of data that has been reported in the literature from trousers fracture tests focuses on fracture propagation. Precisely, the data amounts to a plot of the critical tearing energy $T_c$ as a function of the rate of propagation $\dot{a}(t)$ of the current length of the crack $a(t)$. In truth, it is a plot of
\begin{align*}
\dfrac{2}{B}P(t)\quad {\rm vs.}\quad \dfrac{\dot{l}_0}{2}
\end{align*}
under the assumptions that $T_c= 2P(t)/B$ and $\dot{a}(t)=\dot{l}_0/2$. In the sequel, we present such a plot for the canonical elastomer under investigation here.

\begin{figure}[t!]
  \subfigure[]{
   \label{fig:5b}
   \begin{minipage}[]{0.5\textwidth}
   \centering \includegraphics[width=2.4in]{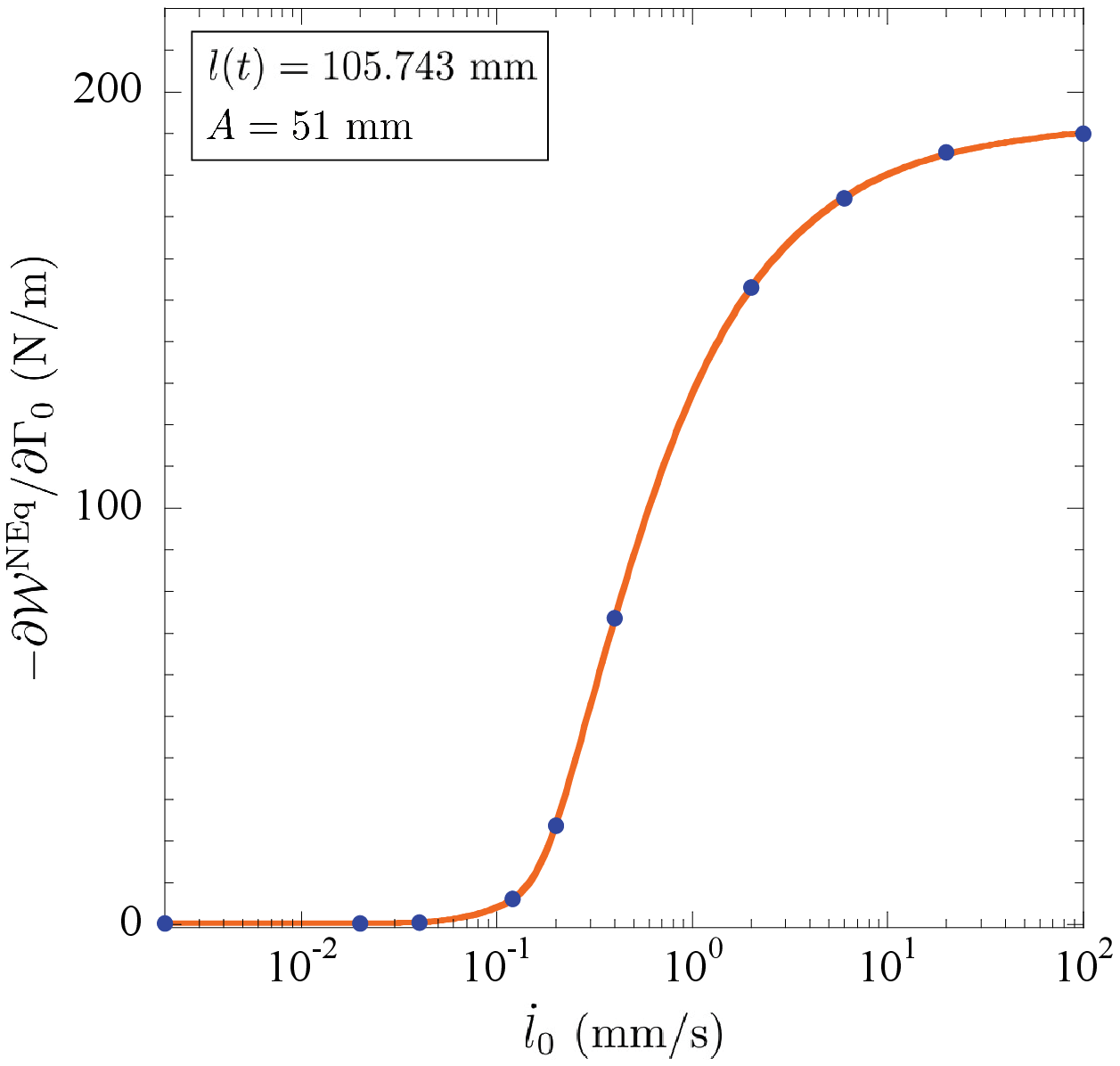}
   \vspace{0.2cm}
   \end{minipage}}
     \subfigure[]{
   \label{fig:5c}
   \begin{minipage}[]{0.5\textwidth}
   \centering \includegraphics[width=2.4in]{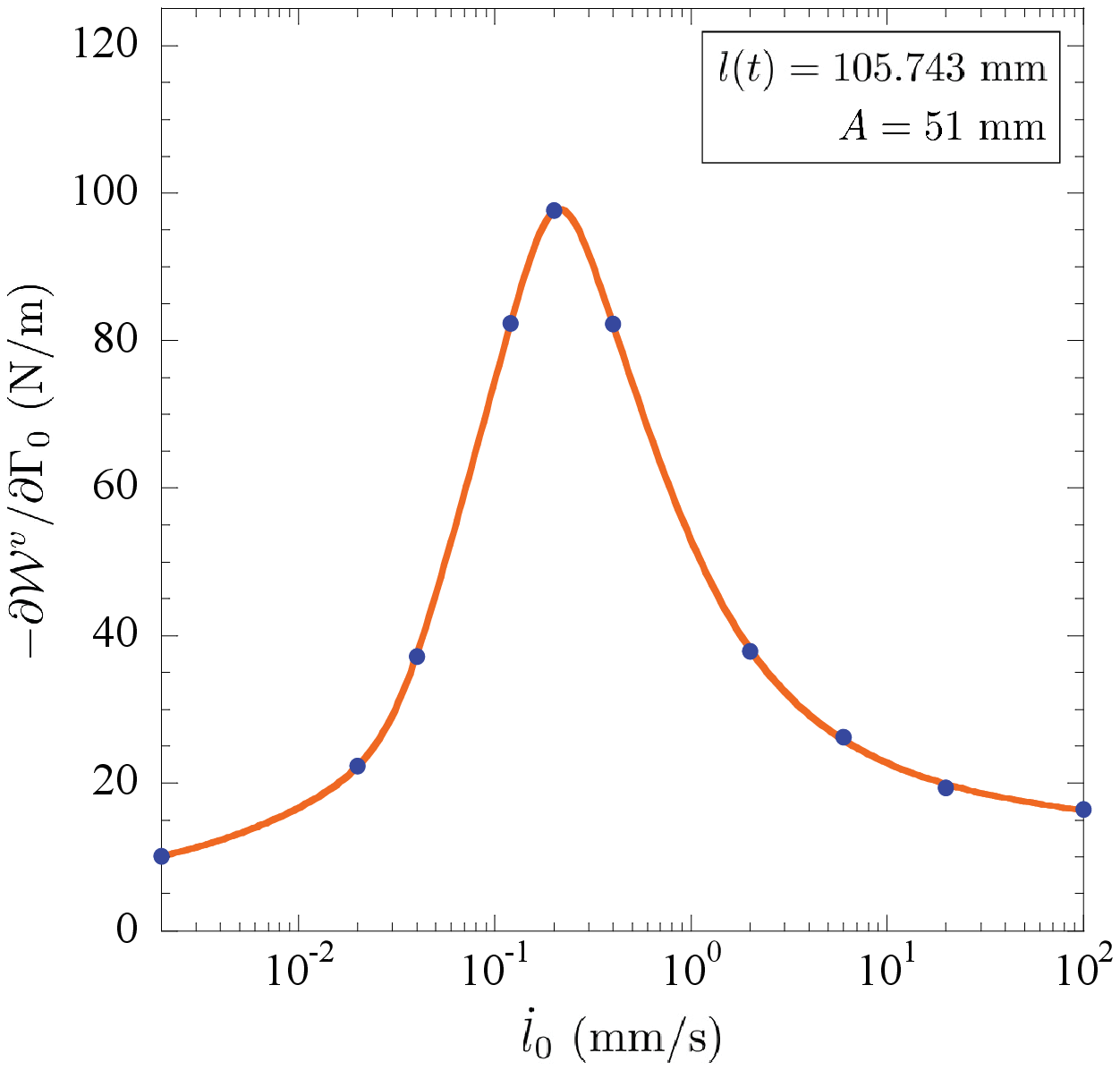}
   \vspace{0.2cm}
   \end{minipage}}
   \caption{Computed values from Figs. \ref{Fig4}(b) and \ref{Fig4}(c) of (a) the derivative $-\partial\mathcal{W}^{{\rm NEq}}/\partial\Gamma_0$ of the non-equilibrium elastic energy and (b) the derivative $-\partial\mathcal{W}^{v}/\partial\Gamma_0$ of the dissipated viscous energy in a trousers specimen with non-equilibrium shear modulus $\nu=2$ MPa, viscosity $\eta=5$ ${\rm MPa}\, {\rm s}$, and a pre-existing crack of length $A=51$ mm deformed at $l(t)=105.743$ mm, plotted as functions of the applied deformation rate $\dot{l}_0$.}\label{Fig7}
\end{figure}

As a first step, note that from the type of 3D plots presented in Figs. \ref{Fig4}(b) and \ref{Fig4}(c) we can compute numerically the derivatives $-\partial\mathcal{W}^{{\rm NEq}}/$$\partial\Gamma_0$ and $-\partial\mathcal{W}^{v}/\partial\Gamma_0$ entering the formula (\ref{Tc-for}) for the critical tearing energy $T_c$. Figure \ref{Fig7} presents the results obtained from this computation in terms of the applied deformation rate $\dot{l}_0$ for a specimen with non-equilibrium shear modulus $\nu=2$ MPa, viscosity $\eta=5$ ${\rm MPa}\, {\rm s}$, and a pre-existing crack of length $A=51$ mm at the same fixed deformation $l(t)=105.743$ mm considered in Figs. \ref{Fig4}(b) and \ref{Fig4}(c).

%

As expected on physical grounds, contrary to the derivative $-\partial\mathcal{W}^{{\rm Eq}}/\partial\Gamma_0$ of the equilibrium elastic energy, note that both  derivatives $-\partial\mathcal{W}^{{\rm NEq}}/\partial\Gamma_0$ and $-\partial\mathcal{W}^{v}/\partial\Gamma_0$ depend strongly on $\dot{l}_0$. In particular, $-\partial\mathcal{W}^{{\rm NEq}}/\partial\Gamma_0$ is bounded from below (by zero) and from above, and increases monotonically with increasing $\dot{l}_0$. By contrast, $-\partial\mathcal{W}^{v}/\partial\Gamma_0$ is also bounded from below (by zero) and from above, but is not monotonically increasing in $\dot{l}_0$, instead, it exhibits a single local maximum at some value of $\dot{l}_0$ (in the present case, around $\dot{l}_0=2\times10^{-1}$ mm/s).

Now, making direct use of the type of results presented in Fig. \ref{Fig7}, we can readily determine the critical tearing energy (\ref{Tc-for}).

\begin{figure}[t!]
   \centering \includegraphics[width=2.5in]{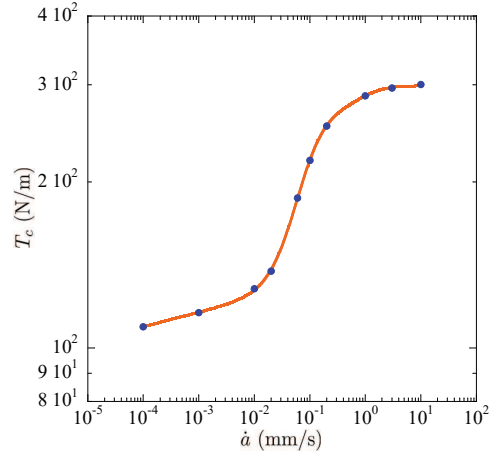}
   \vspace{0.2cm}
   \caption{The critical tearing energy $T_c$ as a function of the rate of crack propagation $\dot{a}(t)$ in a trousers fracture test for a canonical elastomer with intrinsic fracture energy $G_c=100$ N/m, equilibrium shear modulus $\mu=1$ MPa, non-equilibrium shear modulus $\nu=2$ MPa, and viscosity $\eta=5$ ${\rm MPa}\, {\rm s}$.}\label{Fig8}
\end{figure}

To see this via an example, consider once more that the viscoelastic behavior of the elastomer being tested can be described by the canonical behavior assumed in this section, with equilibrium shear modulus $\mu=1$ MPa, non-equilibrium shear modulus $\nu=2$ MPa, and viscosity $\eta=5$ ${\rm MPa}\, {\rm s}$. Consider as well that its intrinsic fracture energy is $G_c=100$ N/m. For this choice of material constants,  as already established above, the Griffith criticality condition (\ref{Gc-0}) is satisfied when the global stretch between the grips is $l_c/l_0=1.0071$. This is precisely the global stretch ($l(t)/l_0=105.743\,{\rm mm}/105\,{\rm mm}=1.0071$) which the results in Fig. \ref{Fig7} pertain to. Then, according to the formula (\ref{Tc-for}), the computation of $T_c$ in this case simply amounts to summing the constant $G_c=100$ N/m to the results in Figs. \ref{Fig7}(a) and \ref{Fig7}(b). Figure \ref{Fig8} reports such a computation of $T_c$ in terms of the rate of crack propagation $\dot{a}(t)$, as defined by $\dot{a}(t)=\dot{l}_0/2$.

%
%

An immediate observation from Fig. \ref{Fig8} is that the critical tearing energy $T_c$ exhibits the ``S'' shape that is the hallmark of trousers fracture tests for viscoelastic elastomers; see, e.g., Fig. 6 in (Greensmith and Thomas, 1955), Fig. 2 in (Mullins, 1959), and Fig. 6 in (Gent, 1996). Specifically, as $\dot{a}\searrow 0$, for sufficiently slow crack propagation rates, $T_c\searrow G_c$. As $\dot{a}$ increases, so does $T_c$ monotonically. As $\dot{a}\nearrow +\infty$, for sufficiently fast crack propagation rates, $T_c$ approaches an asymptotic maximum, $T_{max}$ say. The transition of $T_c$ from its minimum value $G_c$ to its maximum value $T_{max}$ is controlled by both the non-equilibrium elasticity of the elastomer and its viscosity.

\begin{figure}[H]
   \centering \includegraphics[width=2.4in]{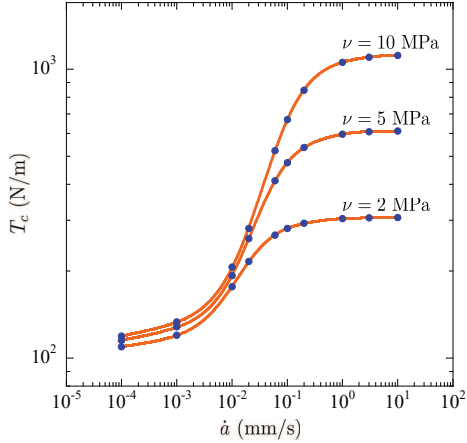}
   \vspace{0.2cm}
   \caption{The critical tearing energy $T_c$ as a function of the rate of crack propagation $\dot{a}(t)$ in a trousers fracture test for a canonical elastomer with intrinsic fracture energy $G_c=100$ N/m, equilibrium shear modulus $\mu=1$ MPa, viscosity $\eta=25$ ${\rm MPa}\, {\rm s}$, and three different non-equilibrium shear moduli $\nu$.}\label{Fig9}
\end{figure}

\paragraph{The effect of the non-equilibrium elasticity} Specifically, for the canonical elastomer under investigation here, the non-equilibrium shear modulus $\nu$ controls the maximum value $T_{max}$ of $T_c$ and, by the same token, how fast $T_c$ increases from $G_c$ to $T_{max}$. Figure \ref{Fig9} illustrates this effect by presenting results of $T_c$ as a function of  $\dot{a}(t)$ for three different non-equilibrium shear moduli, $\nu=2, 5, 10$ ${\rm MPa}$. Save for the value $\eta=25$ ${\rm MPa}\, {\rm s}$ of the viscosity, the values of the remaining material constants are the same as in Fig. \ref{Fig8}.

%
%

\begin{figure}[H]
   \centering \includegraphics[width=2.5in]{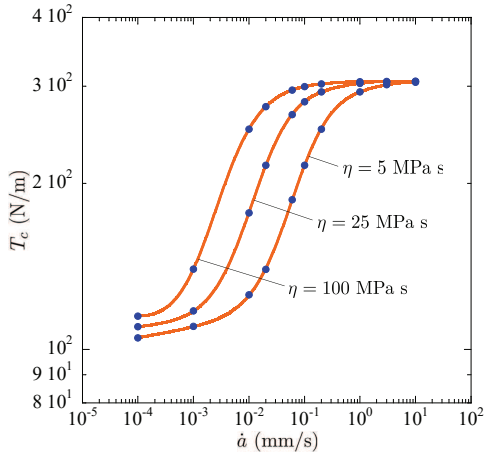}
   \vspace{0.2cm}
   \caption{The critical tearing energy $T_c$ as a function of the rate of crack propagation $\dot{a}(t)$ in a trousers fracture test for a canonical elastomer with intrinsic fracture energy $G_c=100$ N/m, equilibrium shear modulus $\mu=1$ MPa, non-equilibrium shear modulus $\nu=2$ MPa, and three different viscosities $\eta$.}\label{Fig10}
\end{figure}

\paragraph{The effect of the viscosity} On the other hand, for the canonical elastomer under investigation here, the viscosity $\eta$ controls the range of crack propagation rates $\dot{a}(t)$ over which $T_c$ increases from $G_c$ to $T_{max}$. Figure \ref{Fig10} illustrates this effect by presenting results of $T_c$ as a function of  $\dot{a}(t)$ for three different viscosities, $\eta=5, 25, 100$ ${\rm MPa}\, {\rm s}$. The values of the remaining material constants are the same as in Fig. \ref{Fig8}.

%
%

The results in Figs. \ref{Fig9} and \ref{Fig10} make it plain that the critical tearing energy $T_c$, as measured from trousers fracture tests carried out at constant deformation rates $\dot{l}_0$, is primarily a direct manifestation of the viscoelastic behavior --- and \emph{not} of the fracture behavior, as commonly portrayed in the literature --- of the elastomer at hand. More specifically, it is a manifestation of the non-equilibrium elasticity and the viscosity of the elastomer.

\subsection{The local fields in the regions $\emph{\texttt{A}}$, $\emph{\texttt{B}}$, $\emph{\texttt{C}}$, and $\emph{\texttt{D}}$}

\begin{figure}[H]
   \centering \includegraphics[width=3.1in]{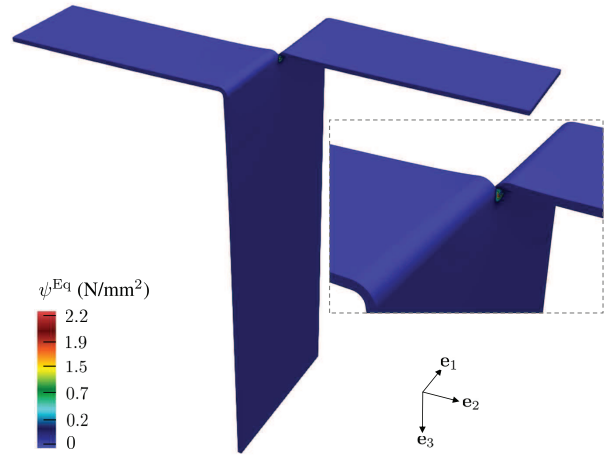}
   \vspace{0.2cm}
   \caption{Contour plot over the deformed configuration of the equilibrium elastic part $\psi^{{\rm Eq}}$ of the free energy (\ref{free-energy}) in a trousers specimen with non-equilibrium shear modulus $\nu=2$ MPa, viscosity $\eta=5$ ${\rm MPa}\, {\rm s}$, and pre-existing crack of length $A=51$ mm deformed at $l(t)=105.743$ mm at a deformation rate $\dot{l}_0=100$ mm/s. The inset provides a close-up of the region around the crack front, where the deformation concentrates. }\label{Fig11}
\end{figure}

For completeness, we close this section by reporting in Fig. \ref{Fig11} a representative contour plot of the equilibrium elastic part $\psi^{{\rm Eq}}(I_1)$ of the free energy (\ref{free-energy}) in a trousers specimen deformed at $l(t)=105.743$ mm at a deformation rate $\dot{l}_0=100$ mm/s. The result pertains to an elastomer with non-equilibrium shear modulus $\nu=2$ MPa, viscosity $\eta=5$ ${\rm MPa}\, {\rm s}$, pre-existing crack of length $A=51$ mm, and is shown over the deformed configuration.

The plot allows to identify the precise locations of the so-called regions $\texttt{A}$, $\texttt{B}$, $\texttt{C}$, and $\texttt{D}$ in the global analysis of the problem; see Fig. \ref{Fig1}. In particular, the plot shows that regions $\texttt{A}$, $\texttt{B}$, and $\texttt{D}$ are substantially undeformed, while the crack-front region $\texttt{C}$ concentrates all the deformation around the crack front.
%
%

\section{Summary and final comments}\label{Sec: Final Comments}

Since the celebrated works of Rivlin and Thomas (1953) and Greensmith and Thomas (1955), experimental studies of nucleation and propagation of fracture from large pre-existing cracks in elastomers subjected to quasi-static mechanical loads have been centered on three types of tests:
\begin{itemize}

\item{the ``pure-shear'' fracture test,}

\item{the delayed fracture test, and}

\item{the trousers fracture test.}

\end{itemize}
In the first two installments of this series --- devoted to deriving the Griffith criticality condition (\ref{Gc-0}) and making use of it to explain these three archetypal fracture tests --- Shrimali and Lopez-Pamies (2023a,b) have explained the ``pure-shear'' and delayed fracture tests. In this paper, the third and final installment, we have made use of the Griffith criticality condition (\ref{Gc-0}) to explain the trousers fracture test.

One of three main results that we have established in this work is that there is a critical global stretch $l_c/l_0$ --- that is, a critical separation between the grips normalized by their initial separation --- at which fracture nucleates from the pre-existing crack in a trousers test, irrespective of the length of the pre-existing crack and of the loading rate at which the test is carried out. The existence of such a critical global stretch appears to have gone unnoticed until now.

Since the early pioneering experiments of Greensmith and Thomas (1955), it has been well documented that in a trousers test carried out at a constant deformation rate $\dot{l}_0$, whenever the crack propagates steadily, the resulting force $P(t)$ at the grips is constant. As a second main result, we have established that this behavior is nothing more than a manifestation of the fact noted above that the Griffith criticality condition (\ref{Gc-0}) in a trousers fracture test happens to be satisfied at a critical global stretch $l_c/l_0$ that is independent of the crack length and of the loading rate.

As a third main result, we have provided quantitative insight into the effects that the non-equilibrium elasticity and the viscosity of the elastomer have on the critical tearing energy $T_c$ obtained from a trousers fracture test carried out at a constant deformation rate. This result makes it clear that $T_c$ is essentially a measure of the capability of the elastomer to dissipate energy through viscous deformation and \emph{not} a measure of its fracture properties. We hope that this result will encourage future experimental studies centered on trousers fracture tests to include separate measurements of the finite viscoelastic behavior of the elastomer being investigated. Regrettably, virtually none of the plethora of experimental studies that have been reported to date in the literature include such measurements.

In conclusion, when viewed collectively, the results presented in this work, together with those presented by Shrimali and Lopez-Pamies (2023a,b) for the ``pure-shear'' and delayed fracture tests, provide broad evidence that the Griffith criticality condition (\ref{Gc-0}) may indeed be the universal condition that governs crack growth in elastomers undergoing finite deformations in response to quasi-static mechanical loads.

Accordingly, as suggested in (Shrimali and Lopez-Pamies, 2023b), given the ``seamless'' mathematical generalization that the Griffith criticality condition (\ref{Gc-0}) provides of the classical Griffith criticality for elastic brittle materials (Griffith, 1921), the next sensible step would be to follow in the footstep of Francfort and Marigo (1998) in order to turn the Griffith criticality condition (\ref{Gc-0}) into a complete mathematical description of the growth of cracks in viscoelastic elastomers.

What is more, as suggested in (Shrimali and Lopez-Pamies, 2023a), it would also behoove us to investigate whether the alluringly simple and intuitive form (\ref{Gc-0}) is in fact universally valid for all dissipative solids, not just viscoelastic elastomers.

\section*{Acknowledgements}

This work was supported by the National Science Foundation through the Grants CMMI--1901583 and CMMI--2132528. This support is gratefully acknowledged.

\end{document}